\documentclass[twoside]{dis07}
\usepackage[latin1]{inputenc}
\usepackage[dvips]{graphicx,epsfig,color}
\usepackage{wrapfig,rotating}
\usepackage{amssymb,amsmath,array}

\begin{document}
\title{Transverse Target-Spin Asymmetry of Exclusive $\rho^0$ Meson Production on Proton at HERMES}

%***********************************************************************
% AUTHORS INFORMATION AREA
%***********************************************************************
\author{Armine Rostomyan$^1$ and Jeroen Dreschler$^2$ \\
%
% DO NOT MODIFY THE FOLLOWING '\vspace' ARGUMENT
(on behalf of the HERMES collaboration) \\
\vspace{.3cm}\\
%
% Addresses and institutions (remove "1- " in case of a single institution)
1- DESY \\
22603 Hamburg, Germany 
%
% Remove the next three lines in case of a single institution
\vspace{.1cm}\\
2- NIKHEF \\
1009 DB Amsterdam, The Netherlands\\
}
%***********************************************************************
% END OF AUTHORS INFORMATION AREA
%***********************************************************************

\maketitle

\begin{abstract}
Preliminary measurements are reported on the azimuthal single-spin asymmetry of exclusive $\rho^0$ mesons for a transversely polarized hydrogen target at HERMES using the $27.6$ GeV HERA positron beam. Within the generalized parton distribution (GPD) formalism, this asymmetry is sensitive to the total angular momentum of quarks and gluons in the nucleon. Since the GPD formalism is only valid for mesons produced by longitudinal photons, the transverse target-spin asymmetry of longitudinal $\rho^0$ mesons is extracted  assuming $s$-channel helicity conservation and compared to theoretical calculations.
\end{abstract}

\section{Introduction}

Hard exclusive meson production in deep inelastic lepton scattering provides access to the unknown generalized parton distributions (GPDs) of the nucleon~\cite{gpd}. For such reactions, it has been shown that for longitudinal virtual photons, the $\gamma^* p $ amplitude can be factorized into a hard lepton-scattering part and two soft parts which parameterize the structure of the nucleon by GPDs and the structure of the  produced meson by distribution amplitude~\cite{fact}. GPDs reflect the 3-dimensional structure of the nucleon and contain information about the total angular momentum carried by partons in the nucleon. Hard exclusive production of $\rho^0$ mesons is sensitive to the GPDs $H$ and $E$ which are the ones related to the total angular momenta $J^q$ and $J^g$ of quarks and gluons in the nucleon~\cite{ji}. The GPD $H$ is already somewhat constrained, while the GPD $E$ is still unknown. In the case of a transversely polarized target, the interference between the GPDs $H$ and $E$ was shown to lead to a transverse target-spin asymmetry (TTSA)~\cite{gpv}. In contrast to the cross section measurements, the TTSA depends linearly on the helicity-flip distribution $E$ with no kinematic suppression of its contribution with respect to the other GPDs. Therefore the TTSA of exclusive $\rho^0$ production can constrain the total angular momenta $J^q$ and $J^g$. 

\section{TTSA of longitudinal $\rho^0$ mesons}

\begin{wrapfigure}{r}{0.45\columnwidth}
%\vspace{-0.9cm}
\centerline{\includegraphics[angle=270, width=4.5cm]{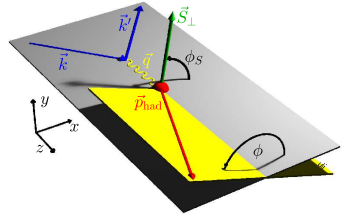}}
\caption{Definition of $\phi$ and $\phi_s$ angles.} \label{Fig:angle}
\end{wrapfigure}

For an unpolarized (U) beam and a transversely (T) polarized target the TTSA $A_{UT}$ is defined as
\begin{equation}
A_{UT}= \frac {1} {P_T} \frac {d\sigma(\phi,\phi_s)-d\sigma(\phi,\phi_s+\pi)} {d\sigma(\phi,\phi_s)+d\sigma(\phi,\phi_s+\pi)},
\end{equation}
where the target polarization $P_T$ is defined w.r.t. the lepton beam direction and the angles $\phi$ and $\phi_s$ are the azimuthal angles of, respectively, the produced $\rho^0$ meson and the target spin vector around the virtual photon direction w.r.t. the lepton scattering plane (see Figure \ref{Fig:angle})~\cite{trento}.
 
The cross section of exclusive $\rho^0$ production can be factorized in terms of angular dependent and angle-independent parts:
\begin{equation}
\frac{d \sigma}{dx_B \, d Q^2 \, dt^\prime \,  d\phi \, d\phi_s } = \frac{1}{4 \pi^2} \frac{d \sigma}{ dx_B \, d Q^2 \, dt^\prime } W(x_B,Q^2,t^\prime,\phi,\phi_s),
\end{equation}
where $x_B$ is the Bjorken scaling variable, $Q^2$ is the squared virtual-photon four-momentum, $t^\prime=t-t_0$. Here $-t$ is the squared four-momentum transfer to the target and $-t_0$ represents the minimum value of $-t$. 

The complete expression for the cross section of $\rho^0$ production is given in~\cite{diehlsap}. The angular distribution $W(\phi,\phi_s)$ can be written\footnote{For simplicity hereafter $x_B$, $Q^2$ and $t^\prime$ are omitted.} in terms of asymmetries:
\begin{equation} \label{eq:wut} 
W(\phi,\phi_s)=\widehat{W}_{UU}(1+A_{UU}(\phi)+P_T A_{UT}(\phi,\phi_s)),
\end{equation}
where $A_{UU}(\phi)=W_{UU}(\phi)/\widehat{W}_{UU}$ is the unpolarized  asymmetry with $ \widehat{W}_{UU} $, $ W_{UU}(\phi) $ being the unpolarized angular distributions and $A_{UT}(\phi,\phi_s)=W_{UT}(\phi,\phi_s)/\widehat{W}_{UU}$ is the transverse asymmetry with the transversely polarized angular distribution $W_{UT}(\phi,\phi_s)$.

Since the factorization theorem is proven for longitudinal photons only~\cite{fact}, the asymmetry of $\rho^0$ mesons induced from longitudinal photons is of theoretical interest. Under the assumption of $s$-channel helicity conservation (SCHC), which implies that a longitudinal vector meson originates from a longitudinal photon, the longitudinal component of the asymmetry is obtained experimentally through the decay angular distribution of $\rho^0$ ($\rho^0 \rightarrow \pi^+ \pi^- $). Each $\rho^0$ helicity state (L, T) results in a characteristic dependence of the  $\gamma^* p$ cross-section on the $\theta_\pi$ polar angle of $\pi^+$ in the $\rho^0$ rest frame~\cite{diehlsap}. The interference terms between different helicities of the $\rho^0$ production are canceled if the cross section is integrated over the $\varphi_\pi$ azimuthal decay angle of $\pi^+$ in the $\rho^0$ rest frame. 

The total angular distribution $W(\cos\theta_\pi,\phi,\phi_s) $, including the dependence on the $\pi^+$ polar angle,  can be written separately for longitudinal $\rho^0_L$ and transverse $\rho^0_T$ mesons:
\begin{eqnarray} \label{eq:wut_sep}
W(\cos\theta,\phi,\phi_s) \,\, & \propto & 
\Big[
\cos^2\theta_\pi \,\, \widehat{W}_{UU}^{\rho_L} \,\, \Big( 1 + A_{UU}^{\rho_L}(\phi) + P_T A_{UT}^{\rho_L}(\phi,\phi_s) \Big) \\ \nonumber
                                   & +       & 
\,\, \,\, \, \sin^2\theta_\pi \,\, \widehat{W}_{UU}^{\rho_T} \,\, \Big( 1 + A_{UU}^{\rho_T}(\phi) + P_T A_{UT}^{\rho_T}(\phi,\phi_s) \Big)
\Big]. \nonumber
\end{eqnarray}   

\section{Extraction of the TTSA}
The data were accumulated with the HERMES forward spectrometer during the running period 2002-2005. The $27.6$ GeV  positron (electron) beam was scattered off a transversely polarized hydrogen target with an average polarization of $0.72$. Events with exactly one positron (electron) and two oppositely charged hadron tracks were selected. Exclusive $\rho^0$ events were identified by requiring $\Delta E=\frac {M_X^2-M_p^2} {2M_p}<0.6$ GeV, where $M_x^2$ is the missing mass squared and $M_p$ is the proton mass. Due to the experimental resolution and limited acceptance, semi-inclusive pion production can contribute to the exclusive sample; this is the primary background. It is well reproduced by the PYTHIA simulation and is estimated to be of the order of $10\%$.

The TTSA asymmetry is extracted by using the unbinned maximum likelihood method where all the moments~\cite{diehlsap} of $A_{UT}(\phi,\phi_s)$, $A_{UT}^{\rho_L}(\phi,\phi_s)$ and $A_{UT}^{\rho_L}(\phi,\phi_s)$ (Eqs. \ref{eq:wut}, \ref{eq:wut_sep}) are fitted simultaneously. In this analysis, the angular distributions $\widehat{W}_{UU}$ and the asymmetries $A_{UU}(\phi)$ of $\rho^0$, $\rho^0_L$ and $\rho^0_T$ meson productions are defined by unpolarized spin density matrix elements (SDMEs)~\cite{schwlf} previously measured by HERMES~\cite{sdme}.

\section{Results}

\begin{wrapfigure}{r}{0.6\columnwidth}
%\begin{figure}
\vspace{-0.6cm}
      \centerline {\includegraphics[width=8.8cm]{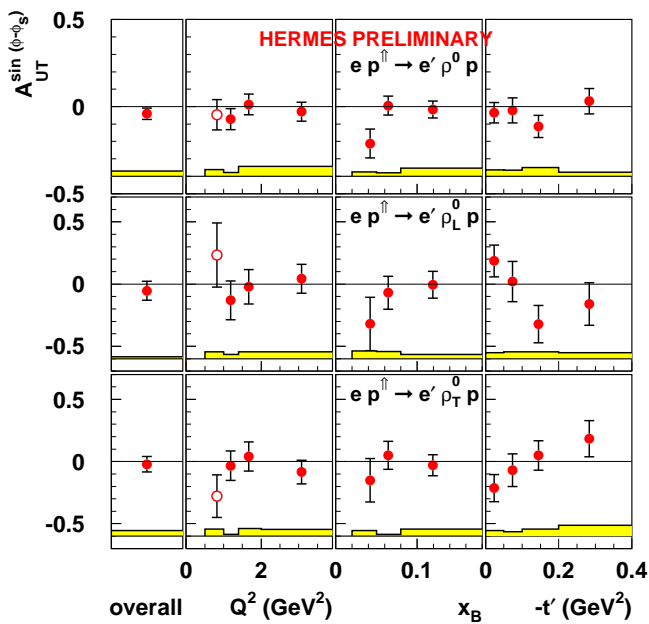}}
%      \centerline {\includegraphics[width=10.8cm]{rostomyan_armine.fig5.eps}}
\vspace{-0.3cm}
      \caption{The integrated value, $Q^2$, $x_B$ and $t^\prime$ dependences of the $A^{\sin(\phi-\phi_s)}_{UT}$ moment of the TTSA of exclusive $\rho^0$, $\rho^0_L$ and $\rho^0_T$ meson productions. }
%The upper panel represents the $\rho^0$ total asymmetries, while the middle and the lower panels represent the longitudinal and transverse asymmetries, respectively.} 
      \label{Fig:a_ut}
%\end{figure}
\vspace{-0.4cm}
   \end{wrapfigure}

The only TTSA moment of $\rho^0$s produced from longitudinal photons that is related to the GPDs $H$ and $E$, is the $\sin(\phi-\phi_s)$ moment. %Under the assumption of SCHC, the $\gamma^*_L$, $\gamma^*_T$ separation is equivalent to a $\rho^0_L$, $\rho^0_T$ separation. 
In Figure \ref{Fig:a_ut} the $A^{\sin(\phi-\phi_s)}_{UT}$ moment of the TTSA is presented. The panels show from left to right the integrated value and the $Q^2$, $x_B$ and $t^\prime$ dependences of the asymmetry. For the $x_B$ and $t^\prime$ dependences, $Q^2$ is required to be above $1$ GeV$^2$. The upper panels represent the $\rho^0$ total asymmetries, while the middle and the lower panels represent the longitudinal $\rho^0_L$ and transverse $\rho^0_T$ separated asymmetries, respectively. The error bars represent the statistical uncertainties only, while the yellow bands indicate the systematic uncertainties due to the target polarization, the background subtraction procedure, the uncertainty resulting from the the unpolarized SDMEs measurement as well as the influence of the beam polarization on the final result.

\begin{figure}
   \hspace{0.5cm}
   \begin{minipage}{5.5cm}
       \centerline {\includegraphics[width=5.5cm]{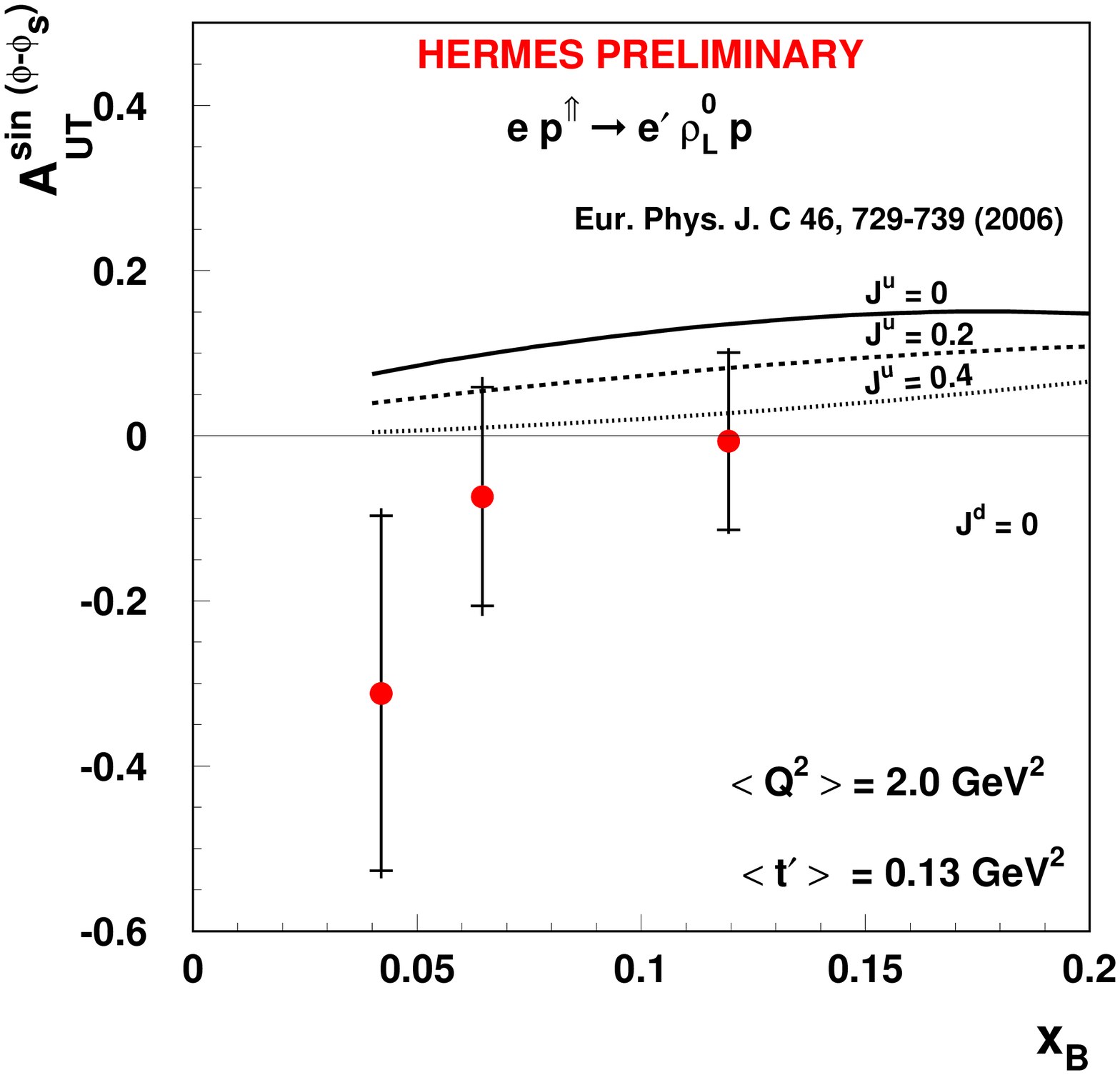}}
   \end{minipage}
   \hspace{1.cm}
   \begin{minipage}{5.5cm}
       \centerline { \includegraphics[width=5.5cm]{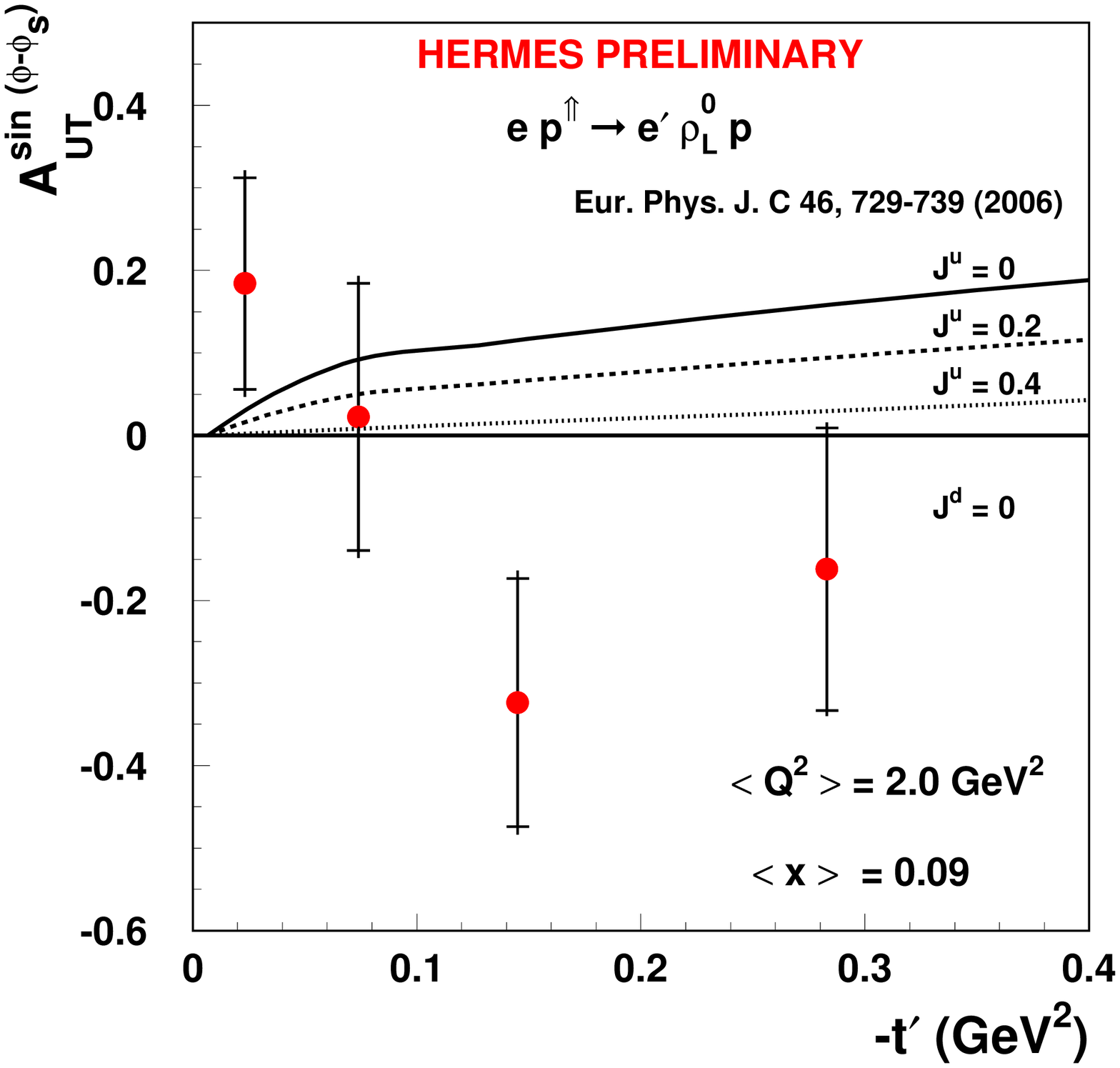}}
   \end{minipage}
   \caption{$x_B$ and $t^\prime$ dependences of $A^{\sin(\phi-\phi_s)}_{UT}$ moment of the TTSA of exclusive production of $\rho^0_L$ mesons compared to the model calculations. The error bars represent the total error.} 
    \label{Fig:a_ut_theor}
\end{figure}

The $x_B$ and $t^\prime$ dependences of the $A^{\sin(\phi-\phi_s)}_{UT}$ moment for longitudinal $\rho^0$ mesons are compared to the theoretical calculations~\cite{vinn} (see Figure \ref{Fig:a_ut_theor}). The longitudinal component of $A^{\sin(\phi-\phi_s)}_{UT}$ moment of the asymmetry is related to: $E/H \propto (E_q+E_g) /(H_q+H_g)$, where the $E_q$, $H_q$ and $E_g$, $H_g$ represent the quark and gluon GPDs, respectively. Currently no model exists for the gluon GPD $E_g$. In the present theoretical calculations the gluon GPD $E$  is neglected. However, $E_g$ is not expected to be large compared to the quark GPDs~\cite{diehl}. No large contribution is expected from sea quarks in our $x_B$ range. As GPD $E_q$ is related to the total angular momentum $J^u$ and $J^d$ carried by $u$ and $d$ quarks, the $A^{\sin(\phi-\phi_s)}_{UT}$ moment of the asymmetry is sensitive to $J^u$ and $J^d$. The various curves in Figure \ref{Fig:a_ut_theor} represent those calculations for $J^u=0$, $0.2$ and $0.4$ and $J^d=0$. The $J^d=0$ choice is motivated by the results of recent lattice calculation~\cite{latt}.
The comparison of $x_B$ and $t^\prime$ dependences of the asymmetry with theoretical calculations indicates that the data favors positive $J^u$ values.

\section{Conclusion}

The $A^{\sin(\phi-\phi_s)}_{UT}$ moment of the TTSA of exclusive $\rho^0$ meson production is measured on a hydrogen target. The kinematic dependences as well as the integrated value of the asymmetry are presented. In particular, the longitudinal part of the asymmetry is compared to theoretical calculations. The model suggests that the data favors positive $J^u$ values, which is in agreement with deeply virtual Compton scattering results obtained from HERMES data~\cite{dvcs}.

\begin{footnotesize}

\end{footnotesize}

\end{document}